\documentclass[lettersize,journal]{IEEEtran}
\IEEEoverridecommandlockouts

\usepackage{mathptmx}  

\usepackage{graphicx}
\graphicspath{{./images/}}

\usepackage{amsmath,amssymb,amsfonts}

\usepackage{algorithmic}
\usepackage{newtxtext,newtxmath}
\usepackage{textcomp}
\usepackage{xcolor}
\usepackage{subcaption}
\usepackage{makecell}
\usepackage{multirow}
\usepackage{hyperref}
\usepackage{stfloats}
\usepackage[numbers,sort&compress]{natbib}  
\usepackage{authblk}
\title{Deep Learning Modeling Method for RF Devices Based on Uniform Noise Training Set}
\author{
Zhaokun~Hu,
Yindong~Xiao,~\IEEEmembership{Member,~IEEE,}
Houjun~Wang,~\IEEEmembership{Member,~IEEE,}
Jiayong~Yu,
Zihang~Gao
}

\date{} 

\begin{document}
\maketitle
\begin{abstract} As the scale and complexity of integrated circuits continue to increase, traditional modeling methods are struggling to address the nonlinear challenges in radio frequency (RF) chips. Deep learning has been increasingly applied to RF device modeling. This paper proposes a deep learning-based modeling method for RF devices using a uniform noise training set, aimed at modeling and fitting the nonlinear characteristics of RF devices. We hypothesize that a uniform noise signal can encompass the full range of characteristics across both frequency and amplitude, and that a deep learning model can effectively capture and learn these features. Based on this hypothesis, the paper designs a complete integrated circuit modeling process based on measured data, including data collection, processing, and neural network training. The proposed method is experimentally validated using the RF amplifier PW210 as a case study. Experimental results show that the uniform noise training set allows the model to capture the nonlinear characteristics of RF devices, and the trained model can predict waveform patterns it has never encountered before. The proposed deep learning-based RF device modeling method, using a uniform noise training set, demonstrates strong generalization capability and excellent training performance, offering high practical application value.\end{abstract}
\section{Introduction}

As integrated circuits (ICs) continue to scale up in size and complexity, modeling radio frequency (RF) chips faces increasingly challenging demands. The primary issues in this domain can be broadly categorized into two aspects: first, the diversity of RF chip types, which makes traditional modeling methods time-consuming and resource-intensive; second, the significant nonlinearities exhibited by RF chips, which traditional techniques struggle to model effectively. To address these challenges, neural networks (NNs) have been introduced as an alternative approach to RF chip modeling. Neural networks are particularly suited to this task due to their broad applicability and ability to learn from relevant data, enabling them to fit a wide range of RF devices. In recent years, deep neural networks (DNNs) have emerged as an effective tool for modeling the nonlinear characteristics of RF devices, thanks to their powerful feature extraction capabilities. As a result, increasing numbers of researchers and engineers have turned to neural networks to overcome the limitations of conventional methods.

IC modeling is generally classified into two primary categories: transistor-level modeling and behavioral-level modeling. Some studies have focused on transistor-level models, where the physical parameters of transistors and operational frequencies are used as inputs to predict performance indicators, such as S-parameters or time-domain outputs, of RF chips using neural networks \cite{guan2021modeling,faraji2023deep,faraji2023new}. However, these models are typically specific to particular chip types and involve complex procedures requiring detailed physical characteristics. In contrast, behavioral-level modeling treats RF integrated circuits as “black boxes,” using neural networks to create input-output models \cite{xu2002neural,oord2016wavenet,dai2017very,liu2023itransformer,liu2004dynamic,charoosaei2023high,nguyen1902fast,pan2023deep,das2023decoder,dhanasekar2023analysis}. While behavioral-level models are generally less accurate than transistor-level models, they offer the advantage of being simpler and faster to construct, making them well-suited for modeling various types of chips. Additionally, behavioral modeling does not require detailed circuit internals, and training and testing data are more readily available. Furthermore, these models tend to exhibit better interpretability. Behavioral-level modeling also reframes the RF modeling task as a sequence feature extraction and prediction problem, where the objective is to create a model that captures the input-output sequence characteristics effectively.

In the context of time-series data prediction, various neural network architectures have been applied to IC modeling, and numerous studies have explored improvements to these architectures. Early research often utilized simpler models, such as Random Forests and Multi-layer Perceptrons (MLPs) for RF chip modeling \cite{xu2002neural,liu2004dynamic} . As research advanced, deeper neural networks were introduced, significantly improving model performance \cite{guan2021modeling,dai2017very} . With deep learning becoming more mainstream, Recurrent Neural Networks (RNNs) emerged as the most widely used time-series prediction model for RF device modeling \cite{faraji2023deep,charoosaei2023high,nguyen1902fast,zebhi2024macromodeling}. Additionally, more advanced architectures, such as Long Short-Term Memory (LSTM) networks \cite{pan2023deep}, Transformers \cite{liu2023itransformer}, AutoEncoders \cite{das2023decoder}, and Residual Networks (ResNets) \cite{dai2017very}, have been shown to be highly effective in time-series prediction tasks. These studies demonstrate that a wide range of neural network structures can be successfully applied to input-output sequence prediction, yielding promising results in RF device modeling.

Despite these advances in neural network architecture, another critical factor influencing the efficiency and comprehensiveness of IC modeling is the dataset used for training. Dataset construction involves two main challenges: the data sources and the waveform features.

The sources of datasets typically include simulation data, public datasets, and measured data. Many researchers rely on simulation waveforms as training data \cite{guan2021modeling,faraji2023deep,charoosaei2023high,nguyen1902fast,faraji2022batch} . However, since simulation conditions often do not fully reflect real-world environments, simulation data may not reliably predict model performance in actual applications. Thus, relying solely on simulated data might not effectively capture the real-world behavior of RF chips. Some researchers have turned to publicly available databases \cite{oord2016wavenet,dai2017very,liu2023itransformer,pan2023deep}, though these datasets typically cover fewer types of integrated circuits and may not meet the diverse modeling and testing needs of all RF devices. In contrast, measured datasets are less commonly used \cite{liu2004dynamic}, primarily because they require specialized hardware and significant experimental effort to construct. However, measured data offers a more accurate reflection of RF chip behavior in real-world environments, making models based on such data more reliable and practical.

In addition to the data source, the waveform features of the dataset are also crucial in determining modeling performance. Common waveform types include sine waves, square waves, audio waveforms, and modulated waveforms. Given the nonlinear nature of RF devices, the performance of these devices can vary significantly under different frequency bands and power conditions. Consequently, to train a neural network capable of learning the full range of an RF device's behavior, the training set should ideally encompass the chip’s performance across various frequencies and input amplitudes. While sine waves cover the full amplitude range, they are limited to a single frequency point in the frequency domain, making them more suitable as a test set rather than a training set. Square waves, with multiple harmonic components, can cover a broader frequency range by adjusting parameters such as rise time and duty cycle, which is why they are often favored as training waveforms \cite{guan2021modeling,faraji2023deep,zebhi2024macromodeling,charoosaei2023high}. However, square waves tend to focus on peak regions in the time domain, which may hinder the model's ability to capture fine-grained time-domain characteristics. To address this limitation, some researchers have turned to more complex waveforms, such as modulated or audio waveforms, for training \cite{oord2016wavenet,dai2017very,liu2004dynamic} , with experimental results showing that these waveforms can help models better extract the chip's characteristics.The suitability of different waveform types for various applications is summarized in Table \ref{tab:waveform_comparison}.
\begin{table*}[ht]
\centering
\caption{Comparison of Waveform Types for Training Sets}
\label{tab:waveform_comparison}
\begin{tabular}{|l|l|l|l|}
\hline
Waveform Type & Frequency Coverage & Amplitude Coverage & Applicable Scenario \\
\hline
Sine Wave & Single frequency point & Full amplitude range & Test Set \\
\hline
Square Wave & Multiple harmonic components & Peak region focused & Training Set (parameter tuning required) \\
\hline
Audio/Modulated Waveforms & Wide frequency band & Complex time-domain features & Training Set (more complex feature extraction) \\
\hline
\textbf{Uniform Noise} & \textbf{Full frequency spectrum} & \textbf{Uniform amplitude distribution} & \textbf{Proposed Training Set for Comprehensive Modeling} \\
\hline
\end{tabular}
\end{table*}

\begin{figure}[ht]
\centering
\includegraphics[width=0.75\linewidth]{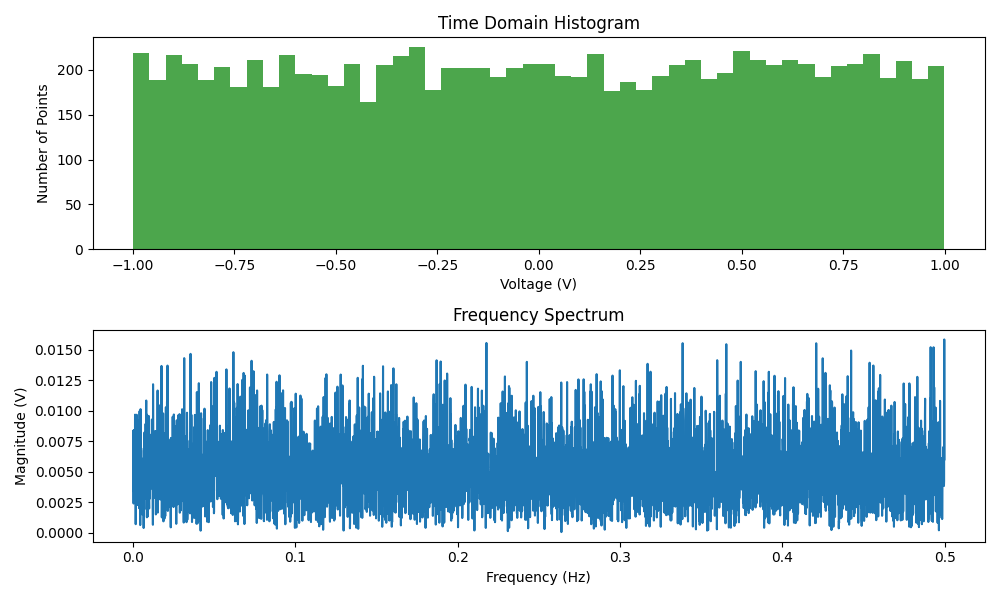}
\caption{Time-domain Histogram and Frequency Spectrum of Uniform Noise}
\label{noise_diagram}
\end{figure}

Previous studies have used waveforms with inherent "information" as training sets. Regardless of the waveform complexity, deep neural networks are capable of learning the underlying chip features. Based on this premise, we propose two hypotheses: first, waveforms that provide broader frequency and amplitude coverage contain more information; and second, neural networks can effectively extract features from even the most complex waveforms. Based on these assumptions, this paper introduces an innovative approach: using uniform noise as the training dataset. The time-domain histogram and frequency spectrum of uniform noise are shown in Figure \ref{noise_diagram}, where the signal is uniformly distributed across both amplitude and frequency domains, theoretically covering all information across various frequencies and amplitudes. If the neural network can extract meaningful features from uniform noise, it will be capable of comprehensively modeling the RF chip's characteristics. 

Building upon this hypothesis, the paper presents a complete IC modeling methodology, which includes system design, data collection, data processing, and model training. The proposed testing system is capable of modeling various RF chips, demonstrating strong versatility. Using an RF amplifier as an example, three deep neural network architectures—AutoEncoder, ResNet, and Mamba—are applied for training and validation. Experimental results show that these models can successfully learn the characteristics of RF chips from uniform noise, and the trained models are capable of predicting the performance of RF devices under different waveforms, validating the effectiveness and universality of the proposed method.

The goal of this work is to design a simple yet effective IC modeling solution that facilitates chip testing and modeling tasks. The paper is organized as follows: Chapter 2 discusses the modeling method, including data collection, dataset generation, data processing, and model training; Chapter 3 presents the experimental setup, where the proposed method is validated using data from the PW210 RF amplifier in both the time and frequency domains; Chapter 4 concludes with a summary of the results, highlighting the effectiveness and advantages of the proposed approach.

\section{Modeling Method Design}

This section provides a detailed description of the RF chip modeling method proposed in this paper, which consists of four key components: data acquisition system setup, dataset generation, dataset processing, and model training conditions.

\subsection{Data Acquisition System}

\begin{figure}[ht]
\centering
\includegraphics[width=0.75\linewidth]{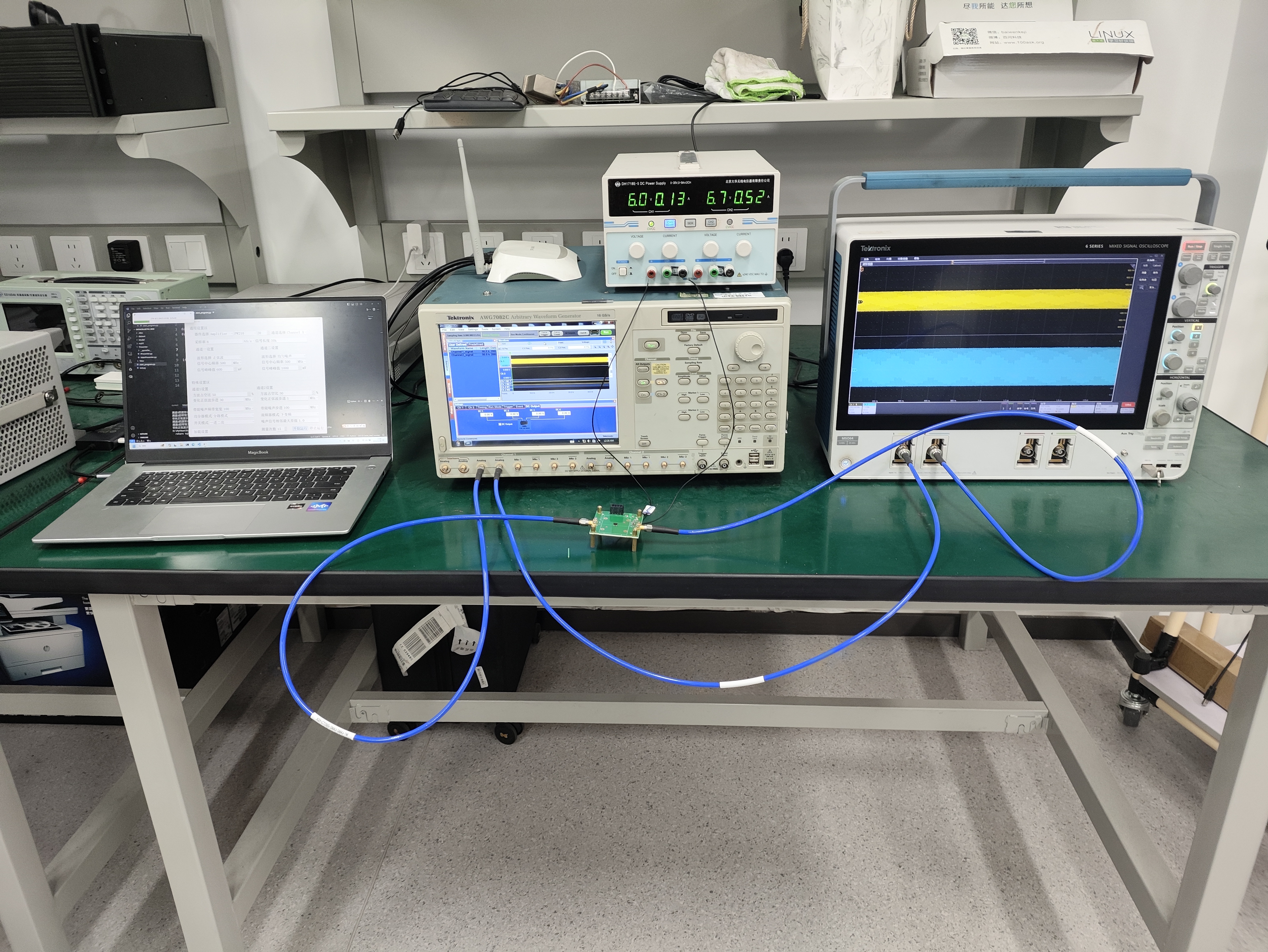}
\caption{Test equipment and their connection relationships}
\label{equipment_real_connection}
\end{figure}

\begin{figure}[ht]
\centering
\includegraphics[width=0.75\linewidth]{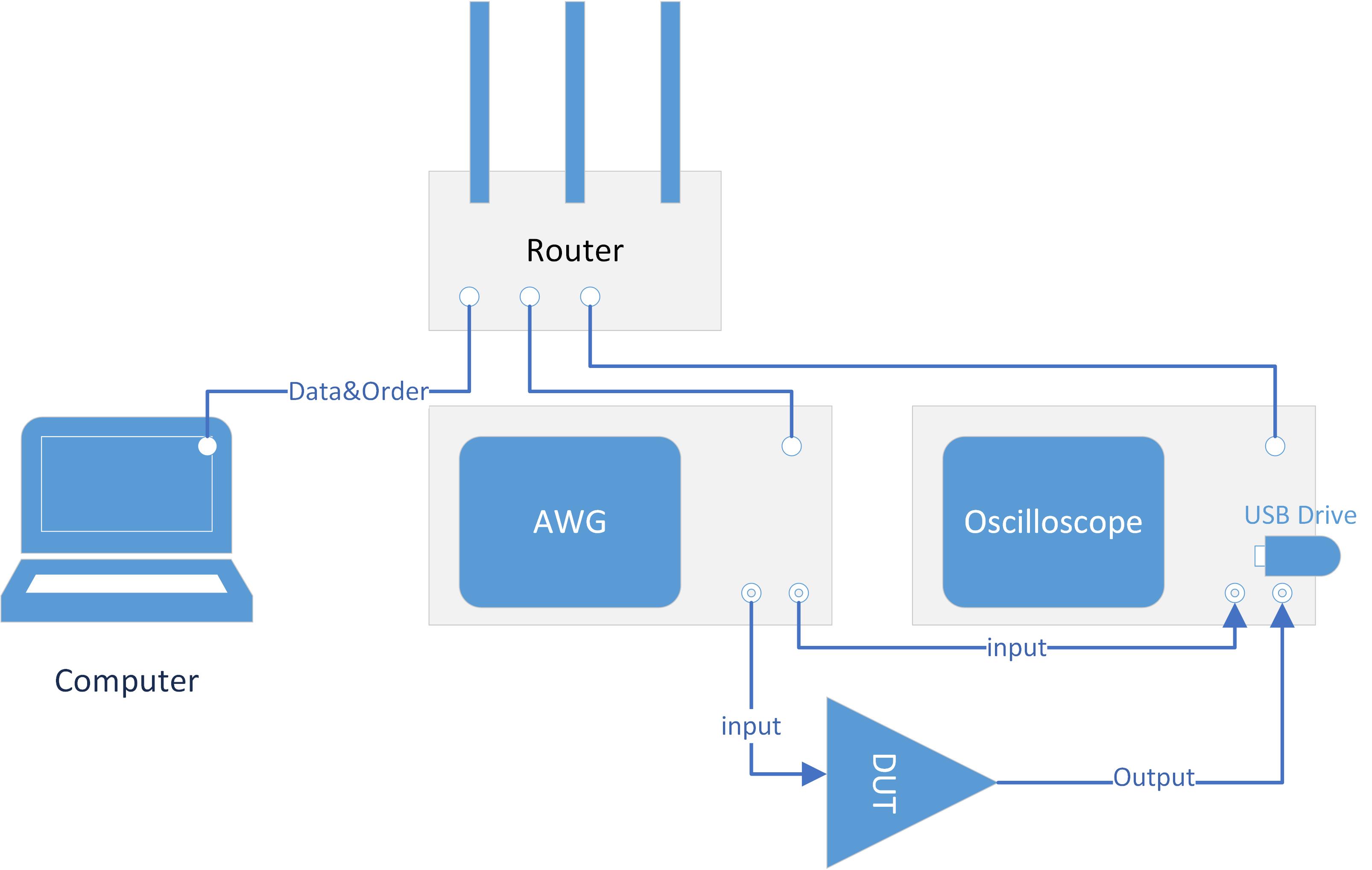}
\caption{Abstract connection diagram of test equipment}
\label{equipment_abstract_connection}
\end{figure}

The collection of time-domain waveform information for RF chip modeling is facilitated by a coordinated setup comprising a waveform generator, an oscilloscope, and a host computer. The hardware connections for the data acquisition system are illustrated in Figure \ref{equipment_real_connection}, with an abstract connection diagram shown in Figure \ref{equipment_abstract_connection}.

The data collection process follows these steps: Initially, the host computer configures the acquisition parameters, including waveform type, number of test points, sampling rate, and the number of acquisitions. Once the configuration is complete, the host computer calculates and generates the digital signals, sending both the waveform data and control commands to the arbitrary waveform generator via a network port. The waveform generator then converts the digital signal into an analog form and outputs it through two ports: one is connected to the input of the chip under test, while the other is connected to one channel of the oscilloscope. The chip's output is connected to the second channel of the oscilloscope. Once both oscilloscope channels have stabilized and received the waveform data, the host computer triggers a single acquisition on the oscilloscope. The collected data is saved in CSV format to an external storage device, such as a USB flash drive. By repeating this procedure, the required number of training and testing datasets can be generated.

\subsection{Dataset Generation}

In order to meet the requirements for various training and test datasets, the host computer must generate different types of signals. The generation methods for each signal type are described below:

\subsubsection{Uniform Noise Signal Generation}

Uniform noise signals are generated by randomly sampling within a specified amplitude range. The mathematical expression for this is:

\begin{equation}
x[n] = \text{Uniform}(-A, A), \quad n = 0, 1, \dots, N-1
\end{equation}

where $x[n]$ is the noise sample generated within the range $[-A, A]$, $A$ is the amplitude of the noise, and $N$ is the number of samples.

\subsubsection{Narrowband Noise Signal Generation}

Narrowband noise signals are created by constructing a complex frequency spectrum within a target frequency band. The spectrum is populated with random values, and frequencies outside the target band are set to zero. The time-domain signal is obtained by performing the inverse Fourier transform of this frequency spectrum, expressed as:

\begin{equation}
X(f) =
\begin{cases}
U(f) + jV(f), & \text{if } f \in [f_\text{start}, f_\text{end}] \\
& \cup [-f_\text{end}, -f_\text{start}] \\
0, & \text{otherwise}
\end{cases}
\end{equation}

\begin{equation}
x(t) = \text{IFFT}(X(f))
\end{equation}

Here, $X(f)$ represents the frequency-domain signal, $U(f)$ and $V(f)$ are the real and imaginary components of the uniform random values, with $U(f), V(f) \sim \text{Uniform}(-1, 1)$, and $f_\text{start}$ and $f_\text{end}$ are the start and end frequencies of the narrowband noise. The time-domain signal $x(t)$ is obtained by applying the inverse Fourier transform to $X(f)$.

The time-domain signal $x(t)$ is normalized to ensure that its amplitude falls within the range $[-1, 1]$, and then scaled according to the target amplitude $A$ using the following equation:

\begin{equation}
x_\text{scaled}(t) = A \cdot \frac{x(t) - \min(x(t))}{\max(x(t)) - \min(x(t))} \cdot 2 - 1
\end{equation}

\subsubsection{Sine Wave Generation}

Sine waves are generated based on a specific frequency, amplitude, and offset. The mathematical expression for a sine wave is:

\begin{equation}
x(t) = A \sin\left(2\pi (f + \Delta f) t\right)
\end{equation}

where $x(t)$ is the amplitude of the sine wave at time $t$, $A$ is the amplitude, $f$ is the reference frequency, $\Delta f$ is the frequency offset, and $t$ is the time, with $t \in [0, T)$ and $T = \frac{N}{f_s}$, where $f_s$ is the sampling rate.

To generate different variations of sine waves, adjustments are made to $A$ for amplitude changes, to $f$ for frequency shifts, and to $\Delta f$ for dual-tone signals, by adding two sine waves with different frequencies.

\subsection{Dataset Processing}

The dataset processing step involves two main operations: delay compensation and dataset normalization, which are essential for improving model accuracy and ensuring proper training.

\subsubsection{Delay Compensation}

Due to transmission line delays, chip delays, and other factors, the output data collected may be delayed relative to the input data. To correct this, cross-correlation is used to adjust for the delay effects. The cross-correlation function $R_{xy}(\tau)$ is calculated as:

\begin{equation}
R_{xy}(\tau) = \sum_{n=-\infty}^{\infty} x[n] \cdot y[n+\tau]
\end{equation}

\begin{figure}[ht]
\centering
\includegraphics[width=0.75\linewidth]{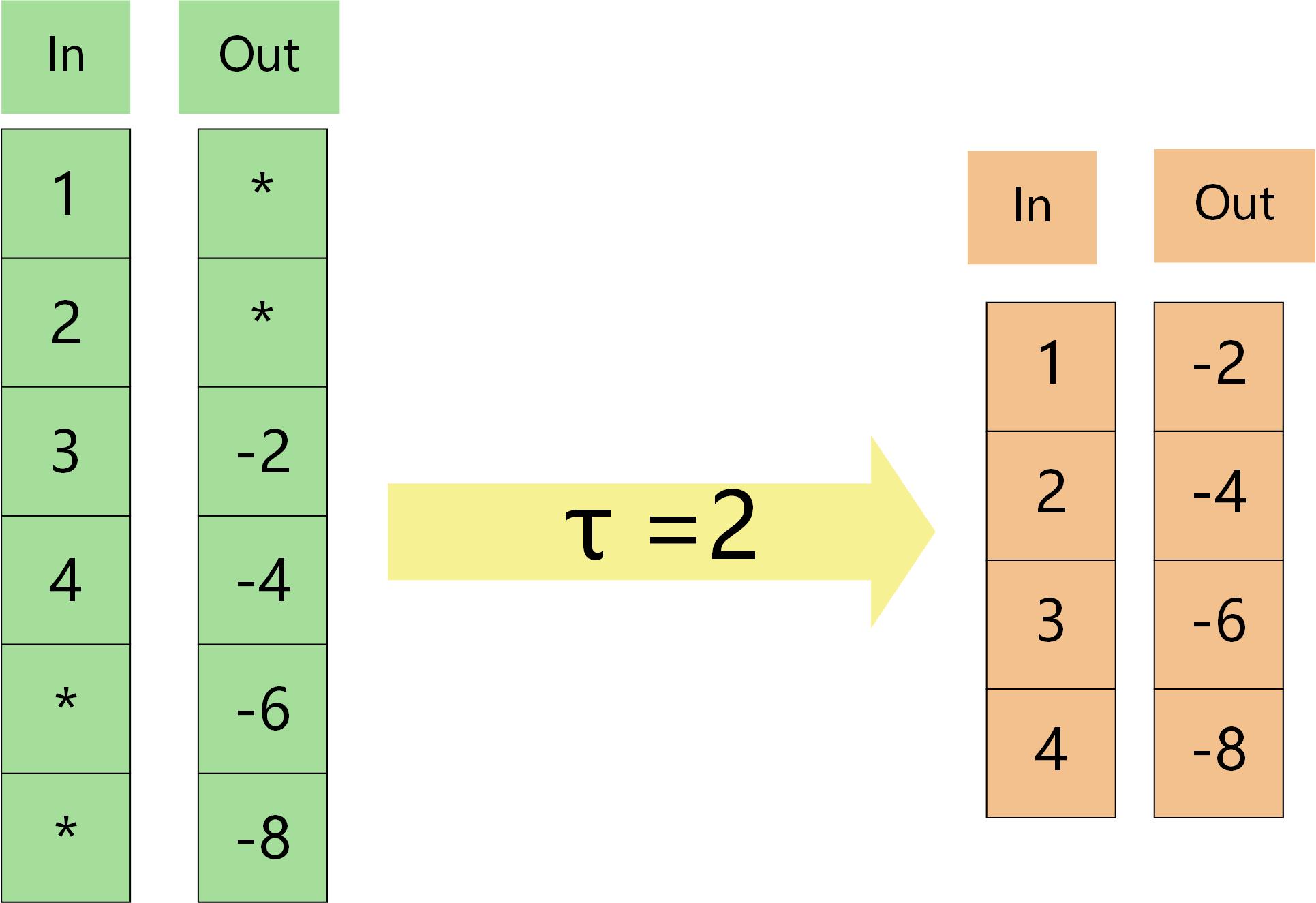}
\caption{Delay correction of sequences}
\label{dealy_remove}
\end{figure}

This function computes the similarity between the input signal $x[n]$ and the output signal $y[n]$ for various delay values $\tau$. By analyzing the cross-correlation for different values of $\tau$, the delay point corresponding to the maximum correlation can be identified. In the case of a reverse amplifier, the delay point corresponds to the minimum negative value of the cross-correlation. Once the delay is identified, the data is corrected, ensuring proper alignment between the input and output signals, as shown in Figure \ref{dealy_remove}.

\subsubsection{Dataset Normalization}

RF integrated circuits are highly nonlinear, and the input-output data ranges can vary significantly between different chips. To address this, normalization is applied to standardize the data. Linear normalization is used to scale the data within the range [0, 1], according to the formula:

\begin{equation}
x' = \frac{x - \min(x)}{\max(x) - \min(x)}
\end{equation}

This normalization ensures that the data is standardized, improving the performance of the neural network models during training.

\subsubsection{Dataset Extraction}

Each waveform file contains approximately 50,000 data points. To prepare these for model training, the data is sliced, and 2048 points are randomly selected as prediction points. The corresponding input data consists of 1024 points taken from the previous time steps. Thus, each test set has the dimensions [2048, 1024, 1], where 1024 points form the input data, which is used to predict the next point in the sequence. By selecting a window size of 1024 points, the model can capture signal features over a longer time range, improving robustness and enabling the extraction of higher-dimensional amplitude and frequency domain characteristics.

\subsection{Model and Training Conditions}

In this study, we selected three typical deep learning network architectures—AutoEncoder, ResNet, and Mamba—to verify the effectiveness and robustness of the proposed method. These three network architectures have been widely discussed and analyzed in existing literature \cite{pan2023deep,he2016deep,gu2023mamba}, and their design and characteristics have been elaborated in detail. Therefore, this paper will not go into redundant descriptions of these networks. The focus of this experiment is to investigate whether these networks can capture and learn the characteristics of the signal contained in uniform noise.

To ensure fairness and comparability of the experiments, we standardized the input window size for the waveform prediction task to 1024 for all three networks. This choice ensures consistency in data input and is also aligned with the typical processing scale used in waveform prediction tasks. Treating these network models as feature extractors, we modified the last layer of each model, adding a fully connected layer to output a single predicted value in order to predict a specific point in the target waveform.

For the loss function, we selected Mean Squared Error (MSE) as the optimization objective, which is a standard loss function widely used in waveform prediction. MSE effectively measures the difference between the predicted and actual values, helping the model gradually optimize prediction accuracy. 
\begin{equation}
\text{MSE} = \frac{1}{N} \sum_{i=1}^{N} (y_i - \hat{y}_i)^2
\end{equation} 
where $y_i$ is the true value, $\hat{y}_i$ is the predicted value, and $N$ is the number of samples.

\begin{figure}[ht] 
  \centering 
  \includegraphics[width=0.75\linewidth]{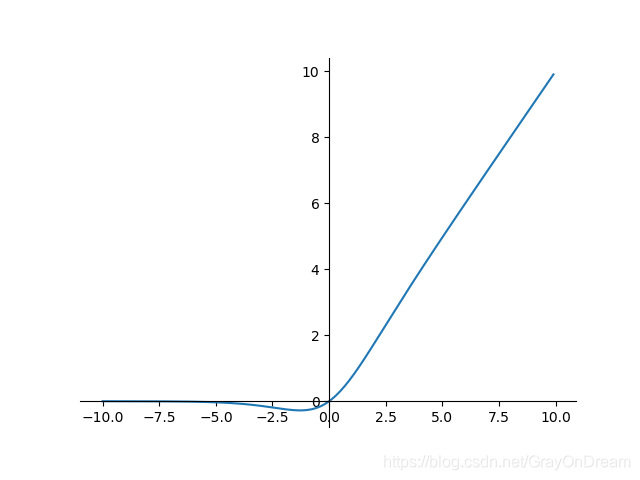} 
  \caption{SiLU Activation Function} 
  \label{collerationPicture} 
\end{figure}

The activation function used in the models is the Sigmoid Weighted Linear Unit (SiLU), which is effective for the nonlinear nature of waveforms, as illustrated in Figure \ref{collerationPicture}. 

The Adam optimizer was used for training. The Adam optimizer combines momentum and adaptive learning rate methods by computing the first and second moment estimates of the gradients, allowing the model to converge more quickly and stably during training.
\begin{equation}
m_t = \beta_1 m_{t-1} + (1 - \beta_1) \nabla_{\theta} J(\theta)
\end{equation}
\begin{equation}
v_t = \beta_2 v_{t-1} + (1 - \beta_2) \nabla_{\theta} J(\theta)^2
\end{equation}
\begin{equation}
\hat{m}_t = \frac{m_t}{1 - \beta_1^t}, \quad \hat{v}_t = \frac{v_t}{1 - \beta_2^t}
\end{equation}
\begin{equation}
\theta_t = \theta_{t-1} - \alpha \frac{\hat{m}_t}{\sqrt{\hat{v}_t} + \epsilon}
\end{equation}
where $\beta_1$ and $\beta_2$ are the decay rates for momentum and RMSProp, $\alpha$ is the learning rate, and $\epsilon$ is a small constant to prevent division by zero errors.

To further improve training efficiency and the model’s generalization ability, we incorporated Batch Normalization (BN) layers into the network architecture. BN standardizes the input to each layer, effectively alleviating the issues of vanishing and exploding gradients during training, accelerating convergence, and thus enhancing overall training performance. The BN formulas are as follows: 
\begin{equation} 
\hat{x} = \frac{x - \mu_B}{\sqrt{\sigma_B^2 + \epsilon}} 
\end{equation} 
\begin{equation} 
y = \gamma \hat{x} + \beta 
\end{equation} 
where $x$ is the input, $\mu_B$ and $\sigma_B^2$ are the mean and variance of the current batch, $\gamma$ and $\beta$ are the learned scaling factor and bias term, and $\epsilon$ is a small constant to prevent division by zero.

In summary, this paper designs and experimentally validates the effectiveness and universality of the three network architectures in the uniform noise learning task.

\section{Experiment and Training Results}

In this experiment, the Device Under Test (DUT) was the PW210 amplifier produced by PreWell. This amplifier was selected due to its distinct nonlinear characteristics, which enable effective testing of the uniform noise training capability and model generalization performance. The signal source used was the Tektronix AWG7082C arbitrary waveform generator, while the Tektronix MSO64 oscilloscope served as the signal monitoring device. Both instruments are programmable, allowing for convenient control through a host computer. 

This section first outlines the dataset design scheme based on this amplifier, followed by validation of the method’s effectiveness and model prediction accuracy from both time-domain and frequency-domain perspectives.

\subsection{Dataset Design}
The test system constructed in the previous section is universal, requiring only adjustments based on the characteristics of the current chip to design the dataset, after which training and testing can proceed. According to the amplifier's datasheet, the maximum input power for the PW210 amplifier is 10 dBm, with a linear operating range from -20 dBm to 5 dBm, and a nonlinear operating range from -5 dBm to 5 dBm. The amplifier’s input impedance is 50 ohms. To convert between amplitude and dBm values, we use the following equation:

\begin{equation} \label{equationdBm} 
P_{\text{in}} = 10 \cdot \log_{10}\left(\frac{2 \cdot V_{\text{max}}^2}{Z \cdot 10^{-3}}\right), \quad \text{dBm}
\end{equation}

where $Z = 50$ ohms and $V_{\text{max}}$ is the amplitude. The factor $2$ in the formula accounts for both the positive and negative half-cycles, while the division by $10^{-3}$ converts power from watts to milliwatts.

Using equation \ref{equationdBm}, we can calculate the following amplitude and dBm relationships:
\begin{itemize}
    \item 10 dBm corresponds to a 2 Vpp sine wave,
    \item -5 dBm corresponds to a 0.25 Vpp sine wave,
    \item 5 dBm corresponds to a 1.12 Vpp sine wave.
\end{itemize}

To cover both the linear and nonlinear regions, the amplitude of the uniform noise is set to 1.2 Vpp, while the amplitude of the band-limited noise in the time-domain test is set to 1 Vpp.

In the frequency domain, the amplifier's three primary parameters are:
\begin{itemize}
    \item Gain-frequency characteristics,
    \item Gain-power characteristics, and
    \item Output Third-Order Intercept Point (OIP3).
\end{itemize}

\begin{figure}[ht] 
  \centering 
  \begin{minipage}{0.45\linewidth}
    \centering
    \includegraphics[width=\linewidth]{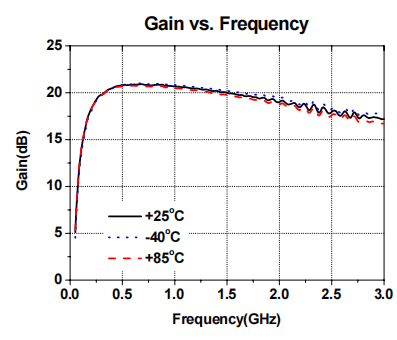}
    \subcaption{Gain-frequency characteristics} \label{fig:gain_frequency}
  \end{minipage} \hfill
  \begin{minipage}{0.45\linewidth}
    \centering
    \includegraphics[width=\linewidth]{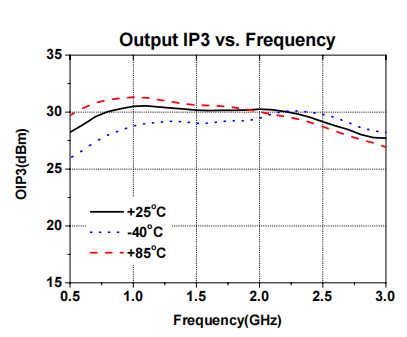}
    \subcaption{Output Third-Order Intercept Point (OIP3)} \label{fig:oip3}
  \end{minipage} \par\vskip\baselineskip
  \begin{minipage}{0.45\linewidth}
    \centering
    \includegraphics[width=\linewidth]{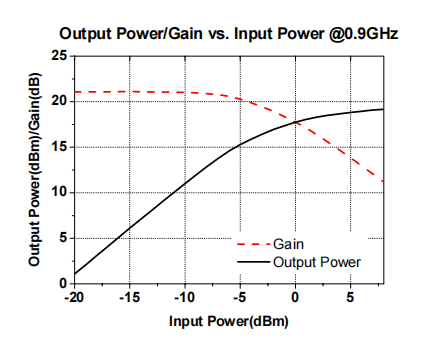}
    \subcaption{Gain-power characteristics (0.9 GHz)} \label{fig:gain_power_0.9G}
  \end{minipage} \hfill
  \begin{minipage}{0.45\linewidth}
    \centering
    \includegraphics[width=\linewidth]{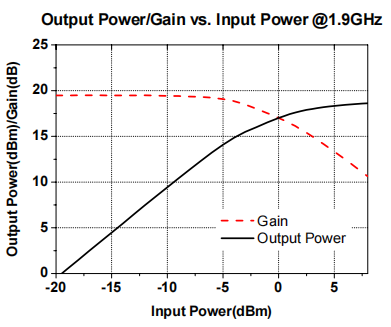}
    \subcaption{Gain-power characteristics (1.9 GHz)} \label{fig:gain_power_1.9G}
  \end{minipage}
  \caption{Frequency-domain parameters from the PW210 datasheet}
  \label{Frequency-domain parameters from the PW210 datasheet}
\end{figure}

These frequency-domain characteristics are shown in Figure \ref{Frequency-domain parameters from the PW210 datasheet}. 

For typical amplifier operation, we refer to the above charts in the datasheet to obtain performance data for the relevant frequency bands. When testing with multiple sine waves of different frequencies and spanning the full frequency range, we can calculate the gain and OIP3 values for each waveform at various frequency points. Connecting these points in the frequency domain should, in theory, recreate the graph presented in the datasheet.

Based on the amplifier's linear operating range, frequency-domain parameters, and other relevant information, the dataset design is summarized in Table \ref{dataset composition}. The reason for using narrow-band noise in the time-domain test set is that narrow-band noise exhibits more complex behavior in the time domain, making it more challenging to predict. This characteristic better demonstrates the model's predictive capabilities. The waveforms used in the frequency-domain test set are designed to assess the model’s performance across the entire frequency range, thereby replicating the frequency-domain parameter charts from the datasheet.

\begin{table*}[ht]
\centering
\caption{Dataset Composition}
\label{dataset composition}
\begin{tabular}{|c|c|c|c|c|}
\hline
\textbf{Type} & \textbf{Waveform} & \textbf{Amplitude} & \textbf{Frequency (Hz)} & \textbf{Quantity} \\ \hline
\textbf{Training Set} & Uniform Noise & 1.2 Vpp & 3 GHz & 300 \\ \hline
\textbf{Time-Domain Test Set} & Band-Limited Noise & 1 Vpp & 0-100 MHz ... 2.9-3 GHz & 30 \\ \hline
\multirow{3}*{\textbf{Frequency-Domain Test Set}} & 
Frequency-modulated sine wave & 0.2 Vpp & 30 MHz, 60 MHz ... 3 GHz & 100 \\ \cline{2-5}
 & Dual-tone sine wave & 0.2 Vpp & 29 MHz + 31 MHz ... 1.999 GHz + 2.001 GHz & 100 \\ \cline{2-5}
 & \multirow{2}*{Amplitude-modulated sine wave} & -20 dBm to -5 dBm & 0.9 GHz & 100 \\ \cline{3-5}
 & & -20 dBm to -5 dBm & 1.9 GHz & 100 \\ \hline
\end{tabular}
\end{table*}

\subsection{Time Domain Training Results}
In this section, three different neural network architectures were trained using time-domain data. The first 200 samples of uniform noise were used as the training set, while the subsequent 50 samples served as the validation set. The batch size was set to 256, and the model was trained for a total of 100 epochs. After training, the model's performance was evaluated on the narrow-band noise test set. The resulting training and testing errors are summarized in Table \ref{Validation and Test Errors}. From the results, it can be observed that all three models fit the training data well and successfully generalized to other types of test sets. Among the tested models, ResNet performed the best in terms of both training and testing error.

\begin{table}[ht]
\centering
\caption{Validation and Test Errors}
\label{Validation and Test Errors}
\begin{tabular}{|c|c|c|}
\hline
\textbf{Model} & \textbf{Validation Error} & \textbf{Test Error} \\
\hline
AutoEncoder & $6.680 \times 10^{-4}$ & $3.243 \times 10^{-4}$ \\
\hline
Mamba & $3.362 \times 10^{-4}$ & $7.992 \times 10^{-4}$ \\
\hline
ResNet & $2.201 \times 10^{-4}$ & $1.836 \times 10^{-4}$ \\
\hline
\end{tabular}
\end{table}

\begin{figure}[ht] 
  \centering 
  \begin{minipage}{0.45\linewidth}
    \centering
    \includegraphics[width=\linewidth]{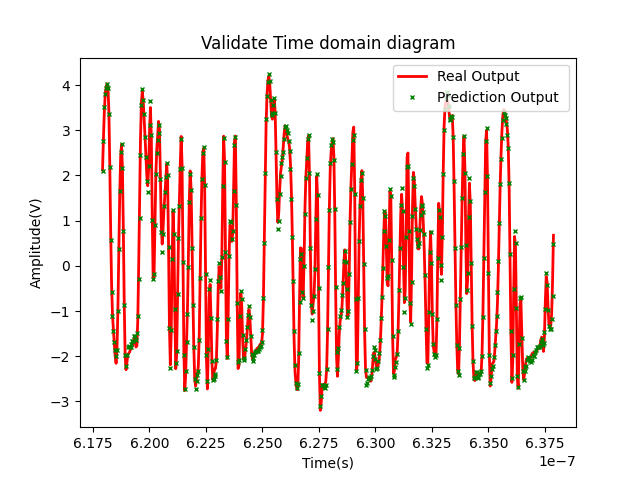}
    \subcaption{Uniform Noise Time Waveform} \label{validate_time_waveform}
  \end{minipage} \hfill
  \begin{minipage}{0.45\linewidth}
    \centering
    \includegraphics[width=\linewidth]{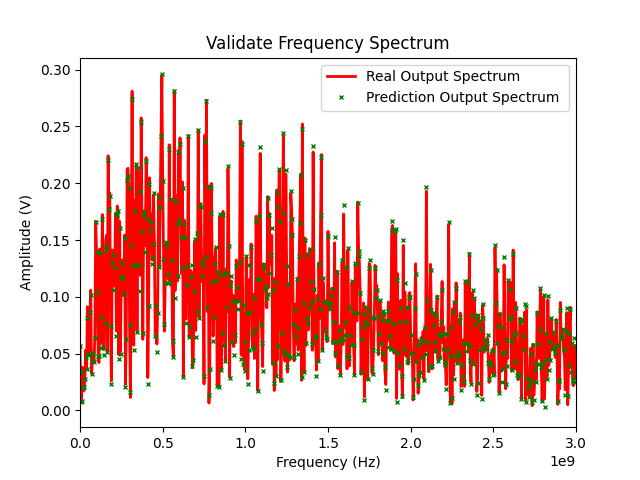}
    \subcaption{Uniform Noise Frequency Spectrum} \label{validate_frequency_spectrum}
  \end{minipage} \par\vskip\baselineskip
  \begin{minipage}{0.45\linewidth}
    \centering
    \includegraphics[width=\linewidth]{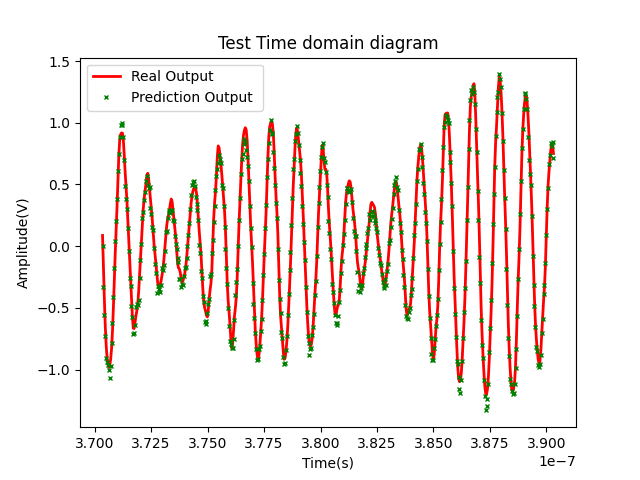}
    \subcaption{Narrow-band Noise Time Waveform} \label{band_noise_time_spectrum}
  \end{minipage} \hfill
  \begin{minipage}{0.45\linewidth}
    \centering
    \includegraphics[width=\linewidth]{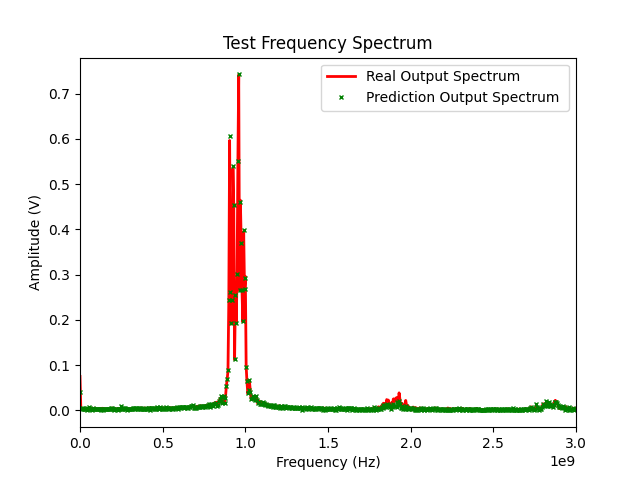}
    \subcaption{Narrow-band Noise Frequency Spectrum} \label{band_noise_frequency_spectrum}
  \end{minipage}
  \caption{Time-domain prediction test}
  \label{time-domain prediction test}
\end{figure}

To further evaluate the performance of the trained ResNet model, it was used for time-domain testing. The task was to predict the uniform noise from the validation set and the band-limited noise from the test set within the frequency range of 900 MHz to 1 GHz. A total of 5000 consecutive points were predicted, with the first 500 points plotted as the time-domain waveform. Subsequently, an FFT was performed on all 5000 points to obtain the frequency-domain spectrum, with the results shown in Figure \ref{time-domain prediction test}.

From the time-domain plot in Figure \ref{validate_time_waveform}, it is clear that the model's predictions are quite accurate in phase, but there is some amplitude error. This is consistent with the frequency spectrum of the uniform noise, which shows a certain degree of attenuation in the high-frequency components. This could be attributed to transmission line losses and the oscilloscope's high-frequency cutoff. In the frequency-domain plot of the band-limited noise (Figure \ref{band_noise_frequency_spectrum}), it is evident that the model accurately predicts the frequency characteristics, capturing even the nonlinear harmonics. This suggests that the model has successfully learned the nonlinear characteristics of the RF device to some extent.

\subsection{Frequency Domain Testing Results}

The frequency-domain parameters of the amplifier mainly include gain and OIP3.

\subsubsection{Gain Parameters}

\begin{figure}[ht]
	\centering
	\includegraphics[width=0.75\linewidth]{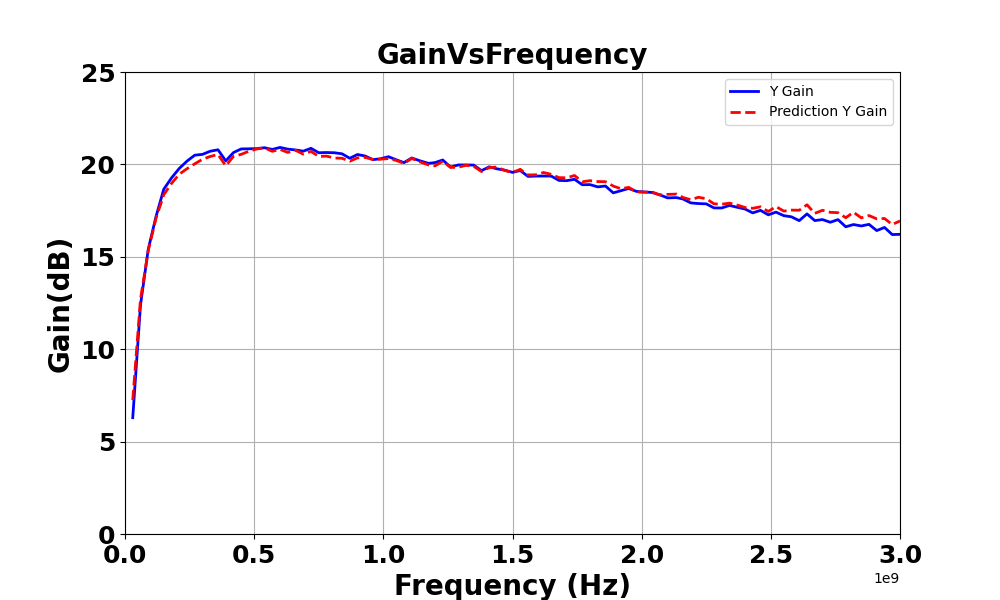}
	\caption{Gain vs Frequency}
	\label{GainVsFrequency}
\end{figure}

The calculation method for gain is as follows:  
First, the trained ResNet model is used to predict 5000 continuous points from 100 sine waves. Afterward, a Fourier Transform (FFT) is applied to the input signal:
\begin{equation}
X(f) = \text{FFT}(\text{Input Signal})
\end{equation}
The frequency with the largest amplitude in the spectrum is identified:
\begin{equation}
f_{\text{max}} = \arg\max_{f} |X(f)|
\end{equation}
The corresponding amplitude is:
\begin{equation}
V_{\text{max}} = |X(f_{\text{max}})|
\end{equation}
The amplitude is converted to power (in dBm) using equation \ref{equationdBm}, denoted as $P_{\text{in}}$.  
The power of the output signal (whether measured or predicted by the model) is computed in the same manner, denoted as $P_{\text{out, measured}}$ and $P_{\text{out, model}}$, respectively. The gain is calculated as:
\begin{equation}
G = P_{\text{out, measured}} - P_{\text{in}}
\end{equation}
and the predicted gain by the model is:
\begin{equation}
G_{\text{model}} = P_{\text{out, model}} - P_{\text{in}}
\end{equation}
For the gain-frequency curve, the frequency and gain at the frequency point with the maximum spectral amplitude are recorded, resulting in 100 gain-frequency points. The gain-frequency curve predicted by the ResNet model is shown in Figure \ref{GainVsFrequency}. It can be seen that the prediction fits well with the measured data up to 2.5 GHz, but there is about a 0.5 dB deviation above 2 GHz.

\begin{figure}[ht] 
  \centering 
  \begin{minipage}{0.45\linewidth}
    \centering
    \includegraphics[width=\linewidth]{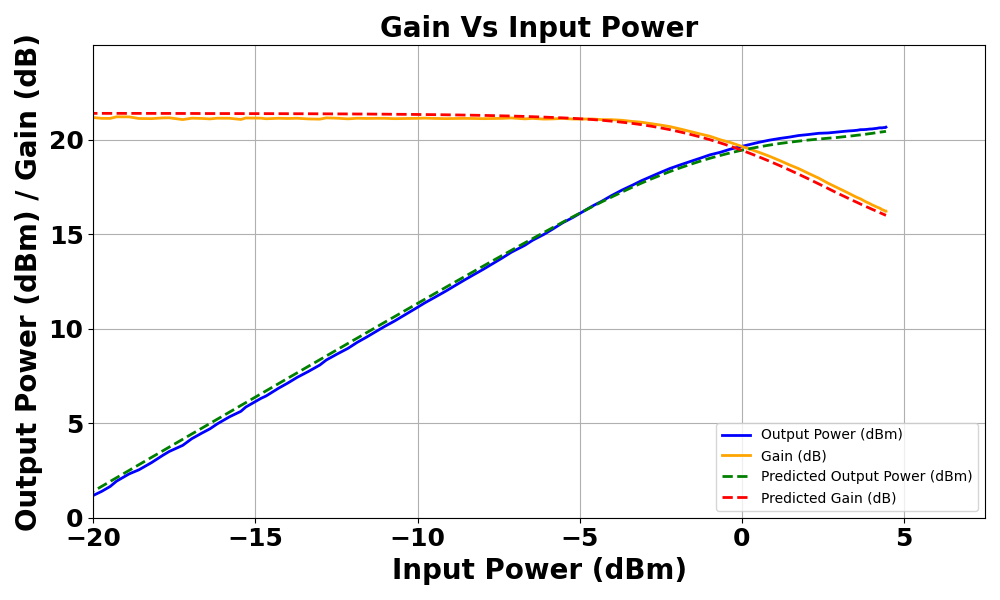}
    \subcaption{Gain vs Input Power at 0.9 GHz} \label{figure_gain_vs_input_power_0.9G}
  \end{minipage} \hfill
  \begin{minipage}{0.45\linewidth}
    \centering
    \includegraphics[width=\linewidth]{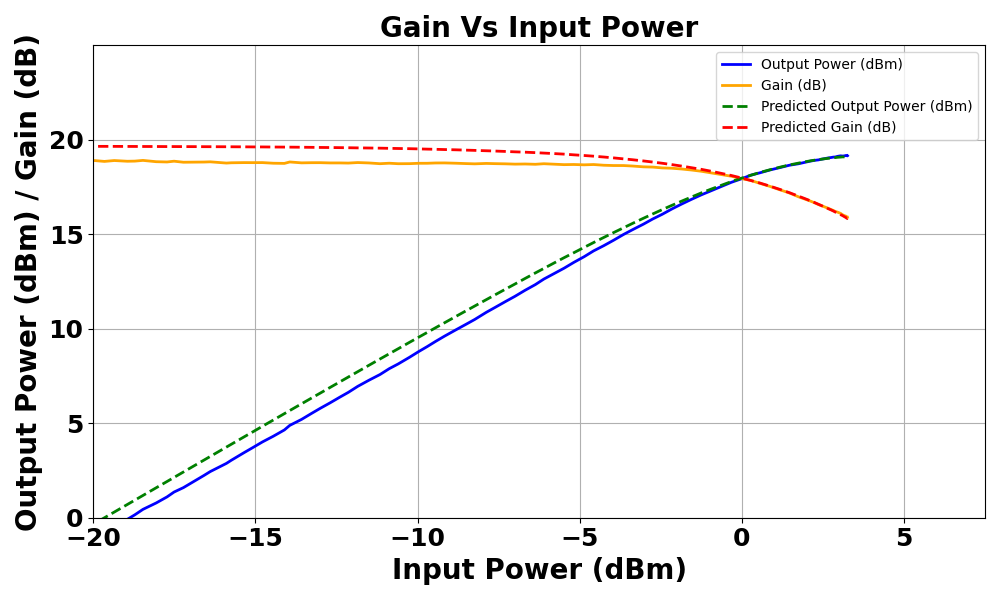}
    \subcaption{Gain vs Input Power at 1.9 GHz} \label{figure_gain_vs_input_power_1.9G}
  \end{minipage} \par\vskip\baselineskip
  \caption{Gain-Input Power Curves}
  \label{figure_gain_vs_input_power}
\end{figure}

For the gain-input power curve, 100 gain-input power points were calculated and plotted. The gain-input power curve predicted by the ResNet model is shown in Figures \ref{figure_gain_vs_input_power_0.9G} and \ref{figure_gain_vs_input_power_1.9G}. At 0.9 GHz, the prediction is quite accurate, but at 1.9 GHz, there is about a 1 dB error at low amplitudes. The error is mainly due to the high-frequency components in Figure \ref{validate_frequency_spectrum}, which are affected by hardware limitations. The model has less training data for higher frequencies, and the amplitude it has learned contains some inaccuracies, leading to this error.

\subsubsection{OIP3}

OIP3 refers to the output power at which the linear portion of the output signal equals the power of the third-order intermodulation product. The calculation principle of OIP3 is shown in the diagram:

\begin{figure}[ht]
	\centering
	\includegraphics[width=0.75\linewidth]{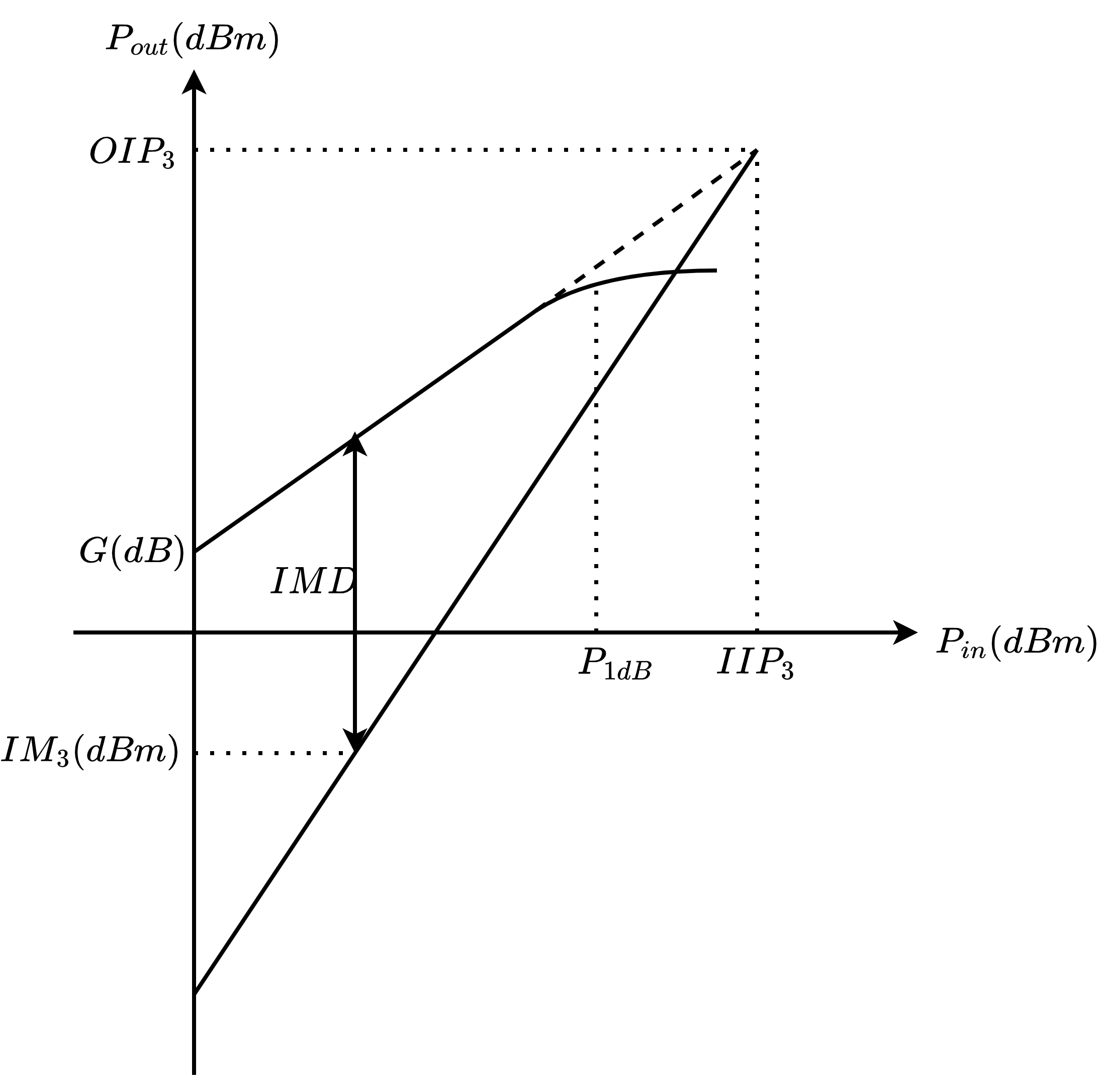}
	\caption{Third-Order Intermodulation Principle}
	\label{third}
\end{figure}

OIP3 cannot be measured directly and must be calculated through derivation. The derivation process is as follows:

1. The first-order output power (linear portion):
\begin{equation}
P_{\text{out,1}} = k_1 + P_{\text{pin}}
\label{equation_Pout1}
\end{equation}
where $k_1$ is the first-order coefficient and $P_{\text{pin}}$ is the input power.

2. The output power of the third-order intermodulation component:
\begin{equation}
P_{\text{out,3}} = k_3 + 3P_{\text{pin}}
\label{equation_Pout3}
\end{equation}
where $k_3$ is the third-order coefficient and $P_{\text{pin}}$ is the input power.

3. By subtracting equations \ref{equation_Pout1} and \ref{equation_Pout3}, we get:
\begin{equation}
3P_{\text{out,1}} - P_{\text{out,3}} = 3k_1 - k_3
\label{equation_subtract_P1P3}
\end{equation}

4. When the first-order output power equals the third-order intermodulation product's output power, i.e., at the third-order intermodulation cutoff point, we have:
\begin{equation}
P_{\text{OIP3}} = \frac{3k_1 - k_3}{2}
\label{equation_OIP3}
\end{equation}

5. Solving these equations gives us the final OIP3 calculation formula:
\begin{equation}
P_{\text{OIP3}} = \frac{3P_{\text{out,1}} - P_{\text{out,3}}}{2}
\end{equation}

Since the frequency difference between the first and third-order components is small (about 2 MHz), additional FFT points are required to distinguish these two frequencies. The formula for calculating the frequency resolution is shown in equation \ref{equation_frequency_resolution}:
\begin{equation}
\Delta f = \frac{f_s}{N}
\label{equation_frequency_resolution}
\end{equation}
where $f_s$ is the sample rate and $N$ is the number of FFT points.
With an oscilloscope sample rate of 25 $GHz$ and FFT points set to 25,000, the frequency resolution is 1 $MHz$, which effectively distinguishes the two signals.

\begin{figure}[ht]
	\centering
	\includegraphics[width=0.75\linewidth]{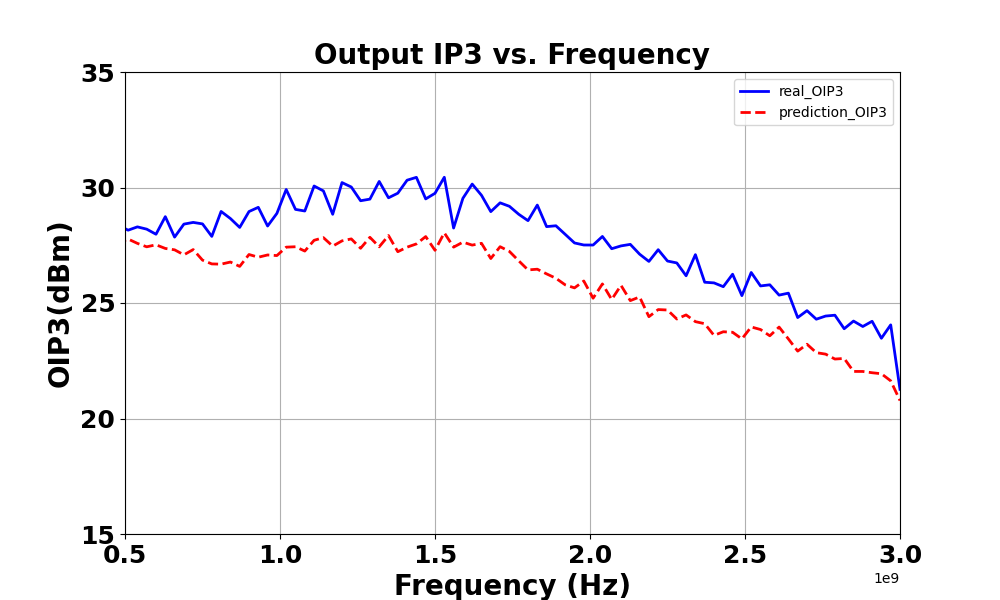}
	\caption{OIP3 vs Frequency}
	\label{figure_OIP3_vs_frequency}
\end{figure}

The OIP3 prediction by the ResNet model is plotted in Figure \ref{figure_OIP3_vs_frequency}. The overall error in the OIP3 prediction is around 2 dBm, primarily due to issues with predicting the third-order intermodulation. The magnitude of the third-order intermodulation is very small, leading to larger errors in dBm. This explains the observed discrepancy in the OIP3 prediction.

\subsection{Summary of Testing Results}

The test results clearly demonstrate that the models trained using the uniform noise training set are highly effective in learning the characteristics of RF devices, with some error margins, in both the time and frequency domains. Notably, the model is also capable of predicting band-limited noise, sine waves, and dual-tone signals, which were not part of the training data. This strongly supports the earlier hypothesis that uniform noise, covering the full amplitude and frequency range of RF device performance, allows neural networks to successfully learn and capture the characteristics of RF devices.

Furthermore, the experiments validate the approach of using multiple time-domain sine waves across the entire frequency range to test frequency-domain performance. This method has proven to be effective within a specific error margin and can be used for comprehensive and detailed model performance testing.

However, there are some limitations with the proposed method. The data collection process is subject to hardware constraints, which introduce certain errors. Despite this, we believe these hardware-related issues can be addressed, and the proposed approach remains practical and feasible.

\section{Conclusion}

This paper presents an innovative deep learning-based approach for modeling RF devices using a uniform noise training set. The effectiveness of this method has been experimentally validated. The model, trained on the noise training set, is capable of accurately predicting various types of signals, showing strong performance in both time and frequency domains.

The key achievements of this paper are:
\begin{enumerate}
    \item A uniform noise training set and corresponding test set are established. The uniform noise training set offers several significant advantages: various neural network architectures can effectively extract features from it; it applies to a wide range of different chipsets; it captures the complete characteristics of RF devices, enabling models trained on uniform noise to predict other previously unseen waveforms; and it is relatively easy to generate and collect.
    \item A comprehensive deep learning-based RF device modeling methodology is designed, covering four essential aspects: data acquisition system setup, dataset generation, data processing, and model training conditions. The dataset design and experimental validation, using the PW210 as an example, were successfully completed. The methodology is generalizable and can be applied to model various RF devices when combined with the uniform noise training set.
\end{enumerate}

The design proposed in this paper successfully addresses two major problems in RF integrated circuit modeling, as discussed at the beginning:
\begin{itemize}
    \item The complexity of different chip types: The solution is adaptable to various chipsets and neural network models.
    \item The challenge of nonlinear fitting: The uniform noise training set captures the full performance characteristics of RF chips, allowing deep learning techniques to model nonlinear behavior effectively.
\end{itemize}

The experiments have thoroughly demonstrated that the proposed method is effective, stable, and reliable, offering substantial practical value.

\bibliographystyle{unsrt}
\bibliography{refs}

\begin{IEEEbiography}
[{\includegraphics[width=1in,height=1.25in,clip,keepaspectratio]{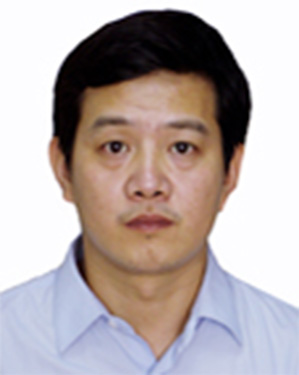}}] 
{Houjun Wang}
	(Member, IEEE) received the M.Sc. and Ph.D. degrees from the University of Electronic Science and Technology of China (UESTC), Chengdu, China, in 1985 and 1992, respectively.
	
	He is currently a Professor and has been a Vice President at Shenzhen Institute for Advanced Study, UESTC, since 2005. His current research interests include time domain measurement and signal processing, design for testability of complex systems, architecture of auto test systems, and fault diagnosis.
\end{IEEEbiography}
\begin{IEEEbiography}
[{\includegraphics[width=1in,height=1.25in,clip,keepaspectratio]{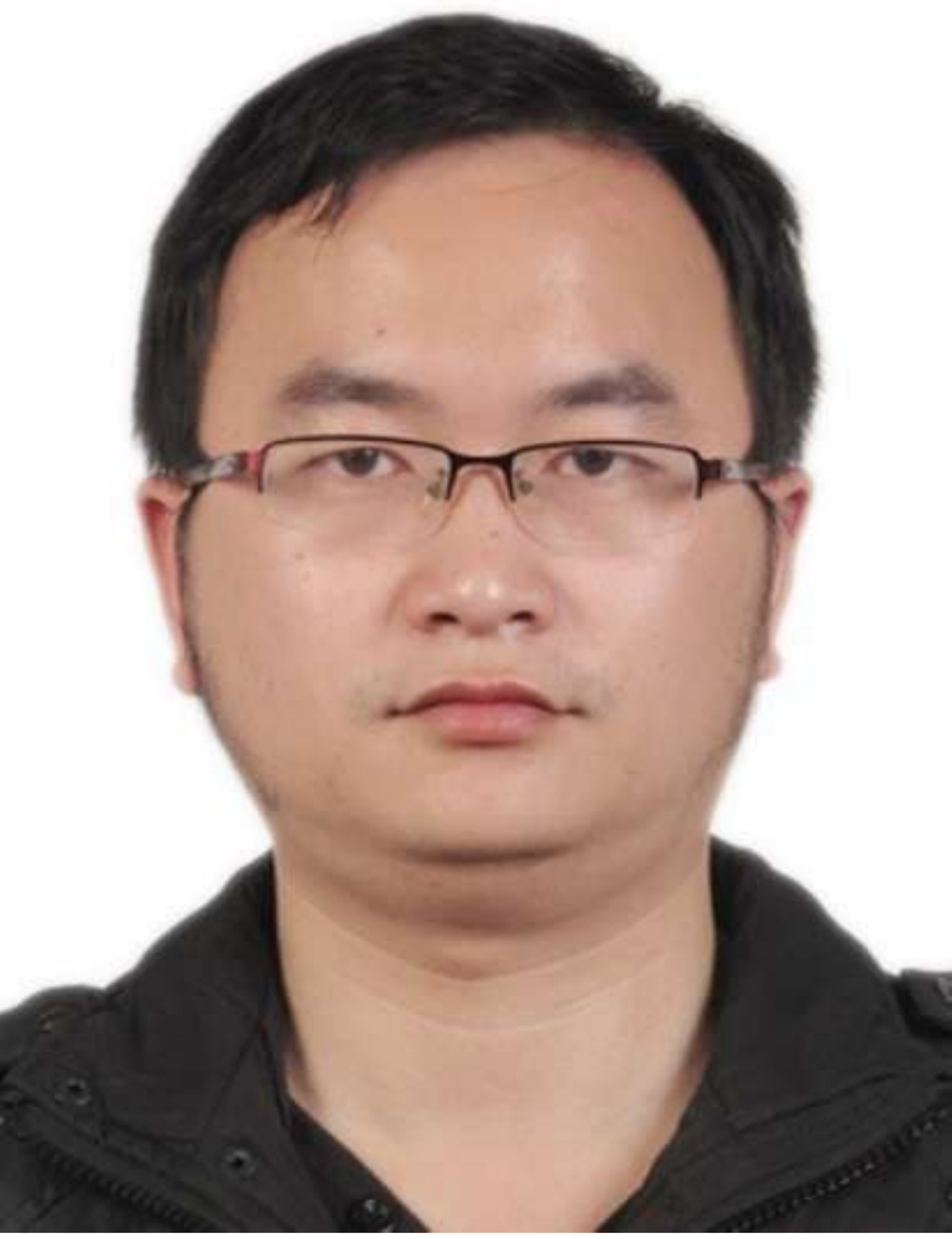}}] 
{Yindong Xiao}
	(Member, IEEE) received the Ph.D. degree from the University of Electronic Science and Technology of China  (UESTC), Chengdu, China, in 2013.
	
	He is currently a Professor at Shenzhen Institute for Advanced Study, UESTC. His current research interests include arbitrary waveform generation techniques, automatic test system architecture, machine learning based test optimization, and computer network testing techniques.
\end{IEEEbiography}
\end{document}